\documentclass[%
 reprint,
 amsmath,amssymb,
 aps,
 pra,
 showpacs
]{revtex4-1}

\usepackage{graphicx}
\usepackage{dcolumn}
\usepackage{bm}
\usepackage{bbold}
\usepackage[export]{adjustbox}
\usepackage{hyperref}
\usepackage{amsmath}
\usepackage{amssymb}
\usepackage{makeidx}
\usepackage{graphicx}
\usepackage{color}

\newcommand{\wrt}{w.r.t.\ }

\newcommand{\hide}[1]{}

\newcommand{\up}[1]{ ^{(#1)}}

\newcommand{\myvec}[1]{\vec{#1}}
\renewcommand{\vr}{{\myvec{r}}}

\newcommand{\vp}{{\myvec{p}}}

\newcommand{\vA}{{\myvec{A}}}

\newcommand{\Om}{\Omega}
\newcommand{\om}{\omega}
\newcommand{\si}{\sigma}

\newcommand{\ep}{\epsilon}

\newcommand{\la}{\lambda}

\newcommand{\Ga}{\Gamma}

\renewcommand{\th}{\theta}

\newcommand{\lcase}{\left\{\begin{array}{ll}}
\newcommand{\rcase}{\end{array}\right.}
\renewcommand{\bar}{\begin{array}{ll}}
\newcommand{\ear}{\end{array}}
\newcommand{\bal}{\begin{align}}
\newcommand{\eal}{\end{align}}
\newcommand{\bma}{\begin{pmatrix}}
\newcommand{\ema}{\end{pmatrix}}
\newcommand{\beq}{\begin{equation}}
\newcommand{\eeq}{\end{equation}}
\newcommand{\bel}[1]{\begin{equation}\label{eq:#1}}
\newcommand{\eel}{\end{equation}}
\newcommand{\bea}{\begin{eqnarray}}
\newcommand{\eea}{\end{eqnarray}}
\newcommand{\beaNN}{\begin{eqnarray*}}
\newcommand{\eeaNN}{\end{eqnarray*}}

\newcounter{lecture}

\newcommand{\Ef}{\mathcal{E}}

\newcommand{\pde}{\partial}
\renewcommand{\hide}[1]{}

\newcommand{\inten}[2]{#1\times10^{#2}\,W/cm^2}
\newcommand{\optc}{\text{ opt.cyc.}}

\begin{document}

\preprint{APS/123-QED}

\title{Electron double-emission spectra for Helium atoms in intense 400 nm laser pulses}

\author{Jinzhen Zhu}
\author{Armin Scrinzi}
\email{Armin.Scrinzi@lmu.de}
\affiliation{%
 Physics Department, Ludwig Maximilians Universit\"at, D-80333 Munich, Germany
}%
\begin{abstract}
Double photoelectron emission from He atoms by intense laser pulses with a wave length of $394.5\,nm$ is computed for 
intensities  $3.5 - 9.2\times 10^{14}W/cm^2$. Joint momentum distributions confirm the characteristics seen
in classical trajectory calculations. 
The pronounced transition from back-to-back to side-by-side emission with increasing intensity, the $He^{++}/He^+$ ratios, and a 
modulation of joint energy spectra agree well with a recent experiment [Henrichs et al., PRA 98, 43405 (2018)], 
if one admits an increase of experimental intensities by a factor $\sim 2$.
We find that Freeman resonances enhance anti-correlated emission, we identify the signature of electron repulsion in 
joint angular distributions, and we interpret the modulation of joint energy spectra as a signature of multiple recollsions.
\end{abstract}

\maketitle

\section{\label{sec:intro}Introduction}
Double-ionization of noble gas atoms has been and still is being investigated for studying the effects of 
elementary correlation and for gauging computational methods. Notably the measurement of enhanced double-ionization
by strong laser pulses~\cite{Walker1994a} has triggered a large number of theoretical studies and consensus
has emerged that  ``recollision'', where the first emitted electron collides 
with the still-bound one, is the primary mechanism of double ionization.
Variants of this basic mechanism have been used to explain in increasing detail spectra using short, intense pulses 
that were obtained by cold target recoil ion momentum spectroscopy (COLTRIMS)~\cite{Ullrich2003}.

The assignment of the observed spectral features to specific mechanisms remains a challenge for theory. 
The Helium atom is, in principle, accessible to a complete numerical solution of its time-dependent
Schr\"odinger equation~(TDSE) and the computation of fully differential spectra, even if the parameter range where
this can be achieved remains narrow. However, an accurate time-dependent wave function by itself does not provide 
physical insight or intuitive mechanisms. For that, the use of classical and semi-classical models is of interest. Such models have been very 
successful in strong field physics~\cite{Katsoulis2018,Vila2018,Price2014}.

The recollision model for non-sequential double ionization (NSDI) consists of  
three steps: 
(1) electron $e_1$ leaves the atom, typically by tunnel ionization; 
(2) $e_1$ picks up energy in the laser field; (3) it returns to the vicinity of the nucleus and collides with the other electron $e_2$. 
The scenarios for the interaction in step (3) are often phrased in terms of classical mechanics. 
If the first electron's energy is large enough, it can knock out the second one in an $e-2e$ collision and the two leave nearly 
at the same time. When the energy imparted to $e_2$ is below the ionization threshold, 
the simultaneous presence of laser can still allow detachment by suppressing the potential barrier. 
The mechanisms where the release occurs within a narrow time-window of the recollision we call ``double-ionization upon recollision''
(DUR), which subsumes direct knock-out and release by suppression of the binding barrier 
as well as tunneling \cite{liu08:helium,haan08:recollision,ye10:doubleionization}. 
 For the correlation of electron momenta, one has to also include scattering of the 
returning electron by the nucleus. Such a process is the "slingshot-NSDI"~\cite{Katsoulis2018}, where the momentum of $e_1$ is
reverted and which leads to the anti-correlated emission of the two electrons.

When an actual excited state is formed with a decay time that is not locked to the recollision event, 
one speaks of ``recollision induced excitation with subsequent ionization'' (RESI)~\cite{Haan2006,Shaaran2010},
which suppresses correlation between the recolliding and the second emitted electron.
A similar pattern, where, however, correlation is maintained, is the formation of a quasi-bound state of both electrons 
which can survive for at least one-quarter cycle and gets ionized with the electrons moving into the same direction 
(``double delayed ejection'',~\cite{Emmanouilidou2011}).

In the present paper we present {\em ab initio} quantum mechanical calculations of double-ionization of the He atom by
short and intense laser pulses at a carrier wavelength of 394.5 nm and relate these to 
recent measurements and some of the mechanisms listed above. Dependence of angular correlation on
pulse intensity and pulse duration is used as the main observable.

We present joint energy and momentum distributions at various intensities and pulse durations and find generally 
good agreement with measurement. Our analysis supports DUR as a main contributor to anti-correlated double emission.
Anti-correlation is further enhanced by Freeman resonances, a genuinely quantum phenomenon. Finally, we will point to 
another manifestation of quantum mechanics, namely the modulation  by $2\hbar \om$ of the joint energy distribution 
along lines of constant sum energy --- the ``checkerboard pattern'' of Ref.~\cite{Henrichs2018a}. 
In classical language this translates into repeated electron collisions.
We also present calculations with ultrashort pulses (2 fs FWHM, parameters of Ref.~\cite{Katsoulis2018}), that
generally support the slingshot mechanism of Ref.~\cite{Katsoulis2018}, although the match is found at lower than predicted 
intensity. 

\section{Methods and laser parameters}
\subsection{Two-electron calculations}
\label{sec:methods}
The Hamiltonian of the He-atom with infinite nuclear mass is (using atomic units $\hbar=e^2=m_e=4\pi\epsilon_0=1$)
\begin{equation}
\label{eq:HamiltonianDI}
 H(\vr_1,\vr_2,t)=H_{I}(\vr_1,t) + H_{I}(\vr_2,t) + \frac{1}{|\vec{r}_1-\vec{r}_2|},
\end{equation}
with the ionic Hamiltonian 
\begin{equation}\label{eq:ionic}
 H_{I}(\vr,t)=-\frac{\Delta }{2}-i\vec{A}(t)\cdot\vec{\triangledown } -\frac{2}{r}.
\end{equation}
Interaction with the laser is described in dipole approximation and velocity gauge, where $\vA$ is defined below.

For numerically solving the TDSE and for computing spectra we use the time-dependent recursive indexing (tRecX) code \cite{Scrinzi}. 
tRecX implements the time-dependent surface flux (tSurff) method~\cite{Scrinzi2012,Tao2012a} (see also 
Refs.~\cite{Hofmann2014,Karamatskou2014,Yue2014a,Majety2015c,Majety2015b,Torlina2015a, Zielinski2016}), 
infinite range exterior complex scaling (irECS)~\cite{McCurdy2004a,Scrinzi2010}, and 
FE-DVR methods~\cite{Rescigno}. In brief, the full two-electron calculation is restricted to 
within a surface radius $|\vr_1|,|\vr_2|\leq R_s$ 
with irECS absorption beyond $R_s$. tSurff is based on the idea that beyond $R_s$ all interactions can be neglected and
spectra are reconstructed from the time-evolution of values and derivatives on a four-dimensional 
hypersurface $|\vr_1|=|\vr_2|=R_s$.
Expansions into single-particle angular momenta and FE-DVR radial functions are used. The most critical convergence parameter is $R_s$ and,
to a lesser degree, the number of angular momenta. All convergence parameters were varied systematically to ensure
sufficient accuracy. In the majority of calculations angular momentum quantum numbers $l_i=0,\ldots,19$ and $|m_i|\leq 1$ were used for 
each electron. Mostly, $R_s=40$ was used, but critical points were investigated with radius $R_s= 80$. We obtain a He ground state energy of 
$-2.902$ with $|m_i|\leq 1$ and the three decimal digits exact value of $-2.903$ with $|m_i|\leq 2$. For a detailed discussion 
of the method and its convergence, see Ref.~\cite{Zielinski2016}. 

Alternative to extracting single emission spectra from the full two-electron calculation, we also used a 
single-active-electron model with the Hamiltonian 
\begin{equation}\label{eq:SIHam}
 H_{M}(t)=-\frac{\Delta }{2}-i\vec{A}(t)\cdot\vec{\triangledown } -\frac{1+e^{-2.135r}}{r},
\end{equation}
which has the ionization potential $I_p=0.903$ a.u. and largely reproduces results from the full calculation, see below.

\subsection{Differential spectra}
Starting from the fully differential momentum spectrum $\si(\vp_1,\vp_2)$ we compute various partially differential spectra.

The co-planar joint angular distributions (JADs) at given energy sharing $\eta=(E_1,E_2)$, $E_i=p_i^2/2m_e$ 
are defined by choosing the first electron at $\th_1$ and taking into account cylindrical symmetry, i.e. 
\beq
JAD(\th_2)=\si(p_1,\th_1,0,p_2,\th_2,\varphi_2)
\eeq
with $\varphi_2=0$ for $\th_2\in[0,\pi]$ and $\varphi_2=\pi$ for $2\pi-\th_2\in[0,\pi]$.
For experimentally realistic JADs we  average over a small energy region $\pm0.3\,eV$, which is comparable to 
the spectral width of the pulses used here. 

Joint distributions of momentum in polarization- ($z$-) direction and joint energy distributions are defined as
\begin{align}
\si(p_{1z},p_{2z})&=\int dp_{1z}dp_{2x}dp_{1y}dp_{1y} \si(\vp_1,\vp_2)\label{eq:specpz}\\
\si(E_1,E_2)&=p_1p_2\int d\Om_1d\Om_2\si(\vp_1,\vp_2),
\end{align}
where $d\Om_i$ is the integration over the solid angle of $\vp_i$.
We adopt the term ``back-to-back'' (B2B) emission for the part of $\si(p_{z1},p_{z2})$ with opposite signs of the $p_{zi}$, 
i.e. the upper left and lower right quadrants in the $(p_{z1},p_{z2})$-plane,
and we call emission with equal signs of  $p_{z1}$ and $p_{z2}$ ``side-by-side'' (SBS).

For the study of (anti-)correlation of double electron emission we introduce the ratio $\Gamma$ of B2B to SBS emission 
\begin{align}\label{eq:gamma}
\Ga&:= Y_{-}/Y_{+},
\\
Y_{\pm}&=\iint_0^\infty\!\! dp_{z1} dp_{z2}[\si(p_{z1},\pm p_{z2}) + \si(-p_{z1},\mp p_{z2})],
\end{align}
such that larger $\Gamma$ indicates more anti-correlation.

We will further study the correlation at individual energy sharing points $\eta=(E_1,E_2)$ using the ratio 
$\Ga_\eta$ where the integration for $Y_\pm$ is restricted to a small region surrounding $p_{iz}=\sqrt{2m_eE_i},i=1,2$.
  
\subsection{Laser pulses}
\label{sec:pulses}

The dipole field of a laser pulse with peak intensity $I=\Ef_0^2/2$ and linear polarization in $z$-direction is defined
as $\Ef_z(t)=\pde_tA_z(t)$ with
\beq
A_z(t)=\frac{\Ef_0}{\om} a(t)\sin(\om t+\varphi_{CEO}).
\eeq
The wave-length was chosen as exactly $\la=394.5\,nm$ to match the experimental wave length used in Ref.~\cite{Henrichs2018a},
with the corresponding photon energy of $\hbar\om\approx3.14\,eV$. For the pulse envelope $a(t)$ we used two different shapes:
a ``flat top'' trapezoidal function with a linear rise and descent over a single optical cycle ($1 \optc=2\pi/\om$) 
and constant amplitude in between.  This somewhat 
unrealistic pulse shape is chosen to better isolate the intensity dependent effects of Freeman resonances. For 
examining the robustness and experimental observability of effects
we chose $a(t)=[\cos(t/T)]^8$ as a more realistic envelope. Pulse durations are specified by the FWHM \wrt intensity.
The carrier-envelope phase $\varphi_{CEO}$, in general, affects all non-linear processes. Even for pulses with a duration of 14\optc
we see some impact on JADs.  With few- or single-cycle pulses, the dependence of spectra on $\varphi_{CEO}$ is very pronounced and
one needs to average over $\varphi_{CEO}$ for comparing to experiments without phase-stabilization.

\subsection{Ponderomotive shifts and Freeman resonances}
The ac-Stark shifts of ground and excited states differ, leading to intensity-dependent resonance conditions known as Freeman resonances~\cite{Freeman1987a}. In good approximation, the shift of excited states energies relative to the ground state is equal to the ponderomotive 
potential $U_p =\Ef_0^2/(4\om^2)$,  leading to the $n$-photon Freeman resonance condition
\begin{equation}\label{eq:freemanCriteria}
-E\up{g} + E\up{x}+U_p=n\omega,
\end{equation}
where $E\up{g}$ and $E\up{x}$ are field-free ground and excited state energies of the He-atom. The validity of this formula for the present purposes was verified
by Floquet calculations with the single-electron Hamiltonian Eq.~(\ref{eq:SIHam}).

Similarly, photo-electron peaks are shifted to lower energies by $U_p$ as the ponderomotive potential of the continuum 
electron is not converted into kinetic energy due to the rapid passage of the pulse. The $n$-photon peaks in single- and double-emission
appear at  energies 
\begin{equation}\label{eq:peaks1}
E\up{1}_n=n\om - I\up{1}_p - U_p
\end{equation}
and
\begin{equation}\label{eq:peaks2}
E\up{2}_n=n\om - I\up{2}_p -2U_p,
\end{equation}
respectively, where $I_p\up{1}$ and  $I_p\up{2}$ are the ionization potentials for single and double ionization.
Note that for the pulse parameters used here, $U_p$ reaches up to several photon energies.

\section{Single electron emission}
In the He atom, single-ionization at longer wave length is little affected by multi-electron effects.
At 800 nm this had been observed for photoemission with linear \cite{Zielinski2016} as well as elliptical
polarization \cite{Majety2015b}. We find the same to hold at the present shorter wavelength.
The difference in total yields obtained from model and full two-electron calculation is about 20\%. After normalization,
the shapes of the spectra agree within a few \% in the energy range up to 100\,eV. As the single ionization calculation
can easily be pushed to complete convergence this also supports the correctness of the full calculation.

\begin{figure}
\includegraphics[width=0.9\columnwidth]{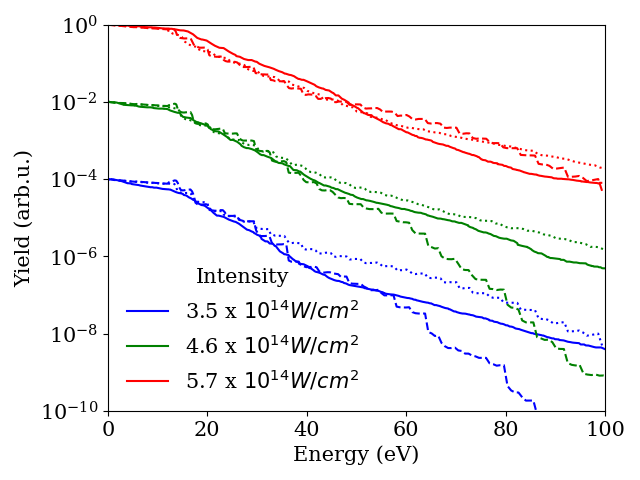}
\caption{
Single-electron energy spectra. Solid lines: experiment~\cite{Henrichs2018a}, dashed: TDSE, dotted: TDSE at
intensities 5.7, 7.4 and $\inten{9.2}{14}$. Curves  were smoothed over 4 photo-electron peaks, normalized, 
and offset artificially for visibility.
The TDSE was solved for a $\cos^8$-pulse with duration 20~fs~FWHM.
}\label{fig:single}
\end{figure}

In Fig~\ref{fig:single} we compare the spectral shapes at two sets of intensities, $3.5,4.6,5.7\times10^{14}$ and $5.7,7.4,9.2\times10^{14}W/cm^2$, respectively, 
to three measured spectra from Ref.~\cite{Henrichs2018a}.
We verified that on the level of the comparison the exact pulse duration does not matter.
The two sets of intensities are chosen \wrt the lowest intensity of $\inten{3.5}{14}$ of Ref.~\cite{Henrichs2018a}: the difference of ponderomotive shifts 
at 3.5 and $\inten{5.7}{14}$ is approximately one photon energy. Photo-electron peaks at the two intensities are 
located at the same energies, just differing by one photon number.
Both triplets of intensities will be used in further comparison with experimental data.

Somewhat surprisingly, for  this rather simple observable the agreement is not satisfactory for either set of intensities. Strikingly, 
at $3.5$ and $4.6\times10^{14}W/cm^2$ the predicted pronounced cutoff is not found in the experimental data. The calculations at the higher set of 
intensities bear more similarity to the experimental data but agreement at the high photo-electron energies remains off by nearly an order of 
magnitude.

The difficulty in using single-electron spectra for intensity calibration is that the photoionization threshold shifts with
intensity by one or several photon energies ($\hbar\om=3.14\,eV$) and channel closure occurs. For example, at intensity 
$\inten{4.6}{14}$ the 10-photon transition falls right onto the ionization threshold and at higher intensity a minimum of 11 photons
is needed for ionization. If the signal is averaged over individual photo-electron peaks, the low-energy photo-electron spectrum
appears to change shape rather erratically. If individual photoelectron peaks were resolved one should be able to reliably gauge the intensity with an ambiguity of multiples of $\hbar\om$. For resolving that ambiguity one needs additional information: the checkerboard pattern observed in double emission (sec.~\ref{sec:checkerboard}) allows distinguishing even and odd photon counts, reducing ambiguity to multiples of two photon energies, $2\hbar\om= 6.3\,eV$. 

The ambiguous comparison of the single-electron spectra precludes the use of these spectra for gauging the experimental intensity. The double emission calculations below suggest that the actual experimental intensities were higher than quoted in \cite{Henrichs2018a}.

\section{Double electron emission}

\subsection{Joint momentum distributions}

\begin{figure}
        \includegraphics[width=\columnwidth, trim=0 0 30 0, clip]{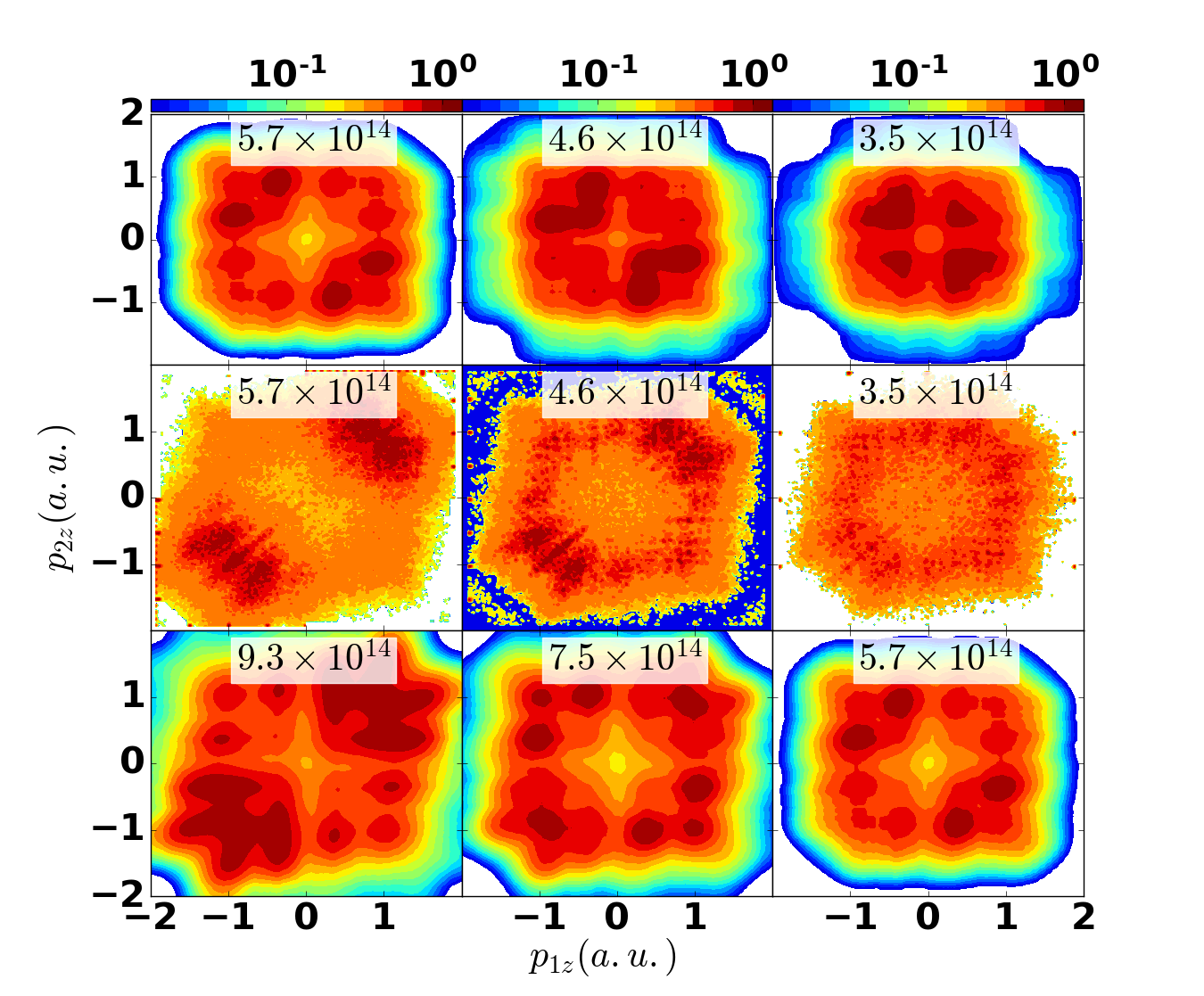}       
    \caption{  \label{fig:doubleSpectra}
    First row: computed spectrum with a 394.5 nm, $cos^8$ pulse. Second row: 
    measured spectra from Ref.~\cite{Henrichs2018a}
    at nominally the same intensities as first row. Third row: computed spectra 
    at a higher set of intensities.
  }
\label{fig:intensityScanningkZ}
\end{figure}

In Fig.~\ref{fig:doubleSpectra} we show the joint momentum distributions obtained at our two intensity sets and the corresponding data 
digitized from Ref.~\cite{Henrichs2018a}.
At the lower intensities  from $3.5$ to $5.7\times10^{14}W/cm^2$ ``back-to-back''(B2B) emission into the quadrants with opposite sign of
the  $p_z$-momentum is more prominent. This changes markedly at $\inten{9.3}{14}$, where the ``side-by-side'' (SBS) emission dominates.
The same transition appears in experiment, although at a nominal intensity near $ \inten{5}{14}$.

We note that the transition to dominantly SBS emission occurs at the intensities in the simulation where the
energy of the recolliding electron approaches the threshold for excitation of $He^+$, cf.~Ref.~\cite{Sheehy1998},
see also Sec.~\ref{sec:antiEnhanScanning}. An inelastic collision at that threshold leaves both electrons at comparatively 
low momentum and unbounded or loosely bound, respectively. From such a state, acceleration by the laser into similar directions is favored.

\subsection{Ratio of $He^{++}$ to $He^+$ yields} 

\begin{figure}[t]
  \includegraphics[width=0.9\columnwidth]{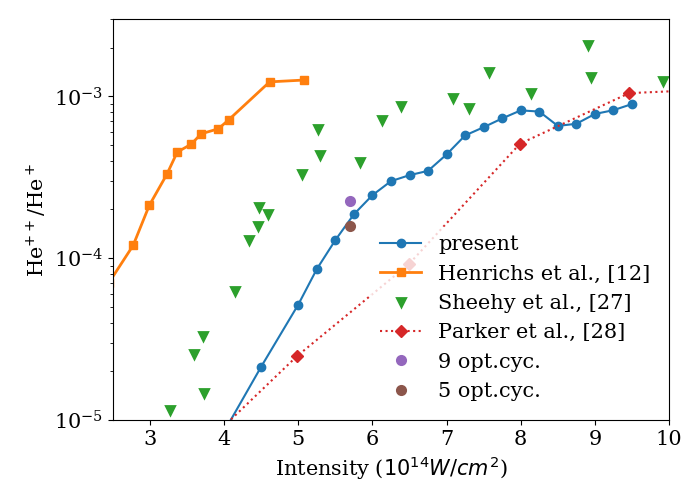}
  \caption{Ratios $He^{++}/He^+$ for $\lambda=390nm$ from present calculation and literature values. Blue line: $7$\,$\optc$ FWHM,
    dots: 5 and $9\optc$ FWHM. 
        }
\label{fig:He2HeRatio}
\end{figure}

The question of experimental intensities also arises, when we consider the ratio $He^{++}/He^+$ of the yields of total double to single ionization. 
Fig.~\ref{fig:He2HeRatio} compares our simulations with the experimental results from references \cite{Sheehy1998} and \cite{Henrichs2018a}. 
Our results suggest that the intensities in both experiments should be scaled to higher values, with about a factor two for 
Ref.~\cite{Henrichs2018a}. The discrepancy to  Ref.~\cite{Sheehy1998} was discussed in
Ref.~\cite{Henrichs2018a} considering in particular the shorter pulse duration used in\cite{Henrichs2018a}. 
For the bulk of our simulations we use short pulses of $\sim 9 fs$ (7 $\optc$  FWHM), even shorter than in Ref.~\cite{Henrichs2018a}. As recollision occurs within one or at most two optical cycles, pulse-duration effects are expected to be small and mostly due to the wider spectrum of shorter pulses. Crosschecks at intensity $\inten{5.7}{14}$ show variations of 
$\sim20\%$ as we change pulse duration from 5 to $9\optc$, see Fig.~\ref{fig:He2HeRatio}. 

Fig.~\ref{fig:He2HeRatio} also includes results from the {\it ab initio} calculation \cite{Parker2000}, which are close to our 
results at most intensities. In Ref.~\cite{Parker2000} yields are accumulated outside a finite radius, which is in 
spirit comparable to the present tSurff calculation, but it differs by the use of flat-top pulses and by the actual extraction method,
which plausibly accounts for the observed differences.

\subsection{The checkerboard pattern}
\label{sec:checkerboard}

\begin{figure}
\includegraphics[width=\columnwidth]{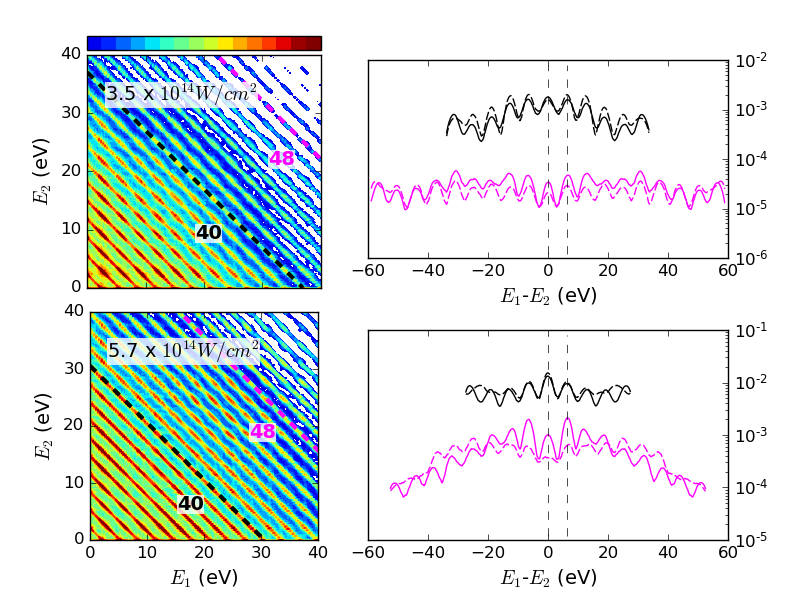}
\caption{\label{fig:checkerboard}
  Modulation of two-electron emission by the photon-energy. Left column: $\si(E_1,E_2)$ for intensities 3.5 and $\inten{5.7}{14}$
  for a flat-top pulse with FWHM=14 \optc. Right column: line-outs at 40 (black) and 48 (magenta) absorbed photons. 
  Solid line is for B2B, dashed is SBS, vertical dashed lines indicate two-photon spacing,
  the $\si(E_1,E_2)$ are normalized to maximum = 1. 
}
\end{figure}

An interesting observation reported in Ref.~\cite{Henrichs2018a} is the appearance of a ``checkerboard'' pattern in the
energy distributions. In Fig.~\ref{fig:checkerboard} we show joint energy spectra at two different intensities and line-outs 
of the spectrum along the 40 and 48-photon peaks according to Eq.~(\ref{eq:peaks2}) for B2B and SBS events separately. 
The line-outs highlight the modulation of the yield at energy differences
$|E_1-E_2|=2n\hbar\om$.  In the line-out for the higher intensity of $\inten{5.7}{14}$ and 48 photons, modulation becomes weaker in the SBS events, but remains pronounced in B2B. These observations are consistent with Ref.~\cite{Henrichs2018a}, where the pattern
was only observed in B2B and became washed out with intensity, although at nominally lower intensities.

In absence of interaction a trivial checkerboard pattern would appear in the emission of two electrons whenever
there are photon-peaks in the emission of the individual electrons. This cannot be the primary cause for the pattern
observed here, as independent (``sequential'') emission of the electrons is several orders of magnitude less intense than 
the recollision induced double emission. In general, periodicity of emission modulates energy patterns at multiples of the 
photon energy, which is interpreted as photon counts and energy conservation, Eq.~(\ref{eq:peaks2}). 
The checkerboard pattern shows that the energy {\em difference} favors multiples of the
photon energy, $E_1-E_2=2n\hbar\om$, which implies multiple interactions between the electrons that are separated 
in time: the mechanism involves 
at least two contributions to double-ionization that are separated by one-half of the optical period. 
Such multiple recollisions where suggested for double-ionization \cite{liu08:helium,haan08:recollision,ye10:doubleionization}, 
being more dominant at lower energies and favoring B2B emission. The energy modulation shown in Fig.~\ref{fig:checkerboard} 
supports these classical predictions. The fact that the pattern appears in experiment in B2B but non in SBS 
emission~\cite{Henrichs2018a} also fits the picture.

\subsection{Anti-correlation and Freeman resonances}\label{sec:antiEnhanScanning}
\begin{figure}
\includegraphics[width=\columnwidth]{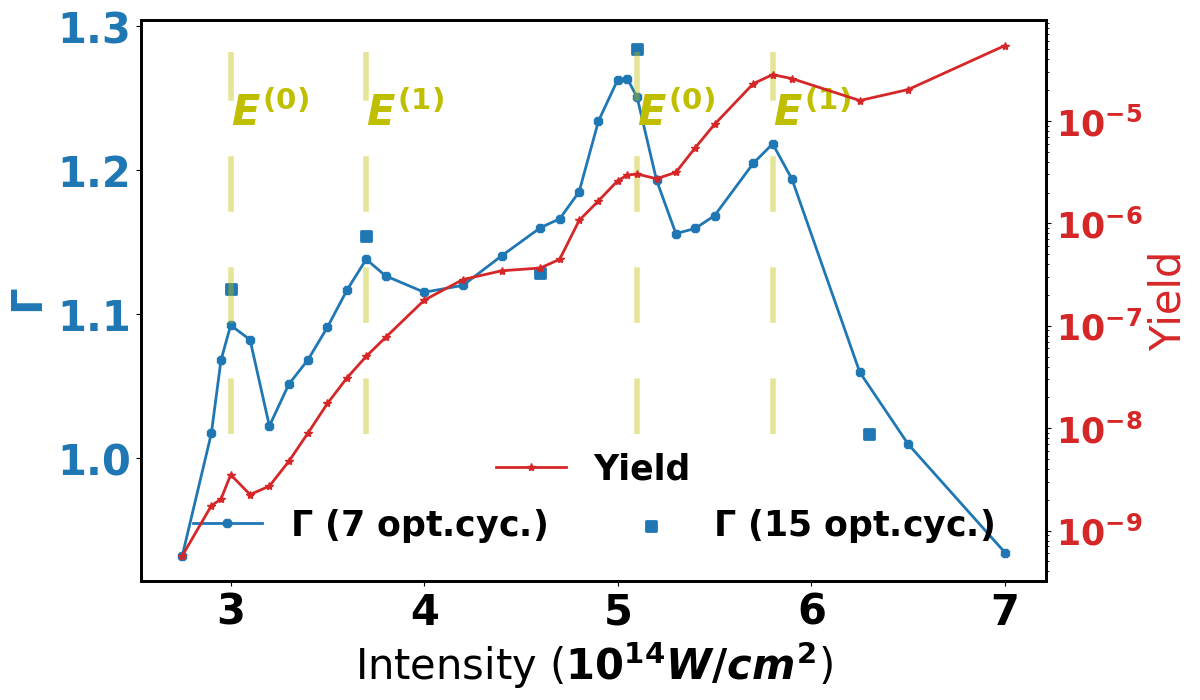}
\caption{Anti-correlation ratio $\Gamma$ and total DI yield as a function of laser intensity.  Blue line: $\Ga$ for 7\optc, dots: 15 \optc, red line: DI yield. 
Dashed lines indicate the Freeman resonances for the two lowest excited states. A $9\optc$ flat-top pulse was used.
}
\label{fig:intensityScanning}
\end{figure}

Fig.~\ref{fig:intensityScanning} shows the anti-correlation ratio $\Ga$, Eq.~(\ref{eq:gamma}), and the total double-ionization yields for intensities from 
$2.5$ to $\inten{7}{14}$. In both curves we see peaks when lowest excited energies $E\up{x}$ shift into Freeman resonance, 
Eq.~\ref{eq:freemanCriteria}. The curves are calculated with a 9 $\optc$  $\sim 12\,fs$ flat top pulse. 
A few additional points were calculated with a pulse duration of 15 $\optc$:  $\Ga$ is further enhanced and while  it drops slightly off-resonance, as to be expected.
\begin{figure}
    \includegraphics[width=0.49\columnwidth, trim=25 25 25 25, clip]{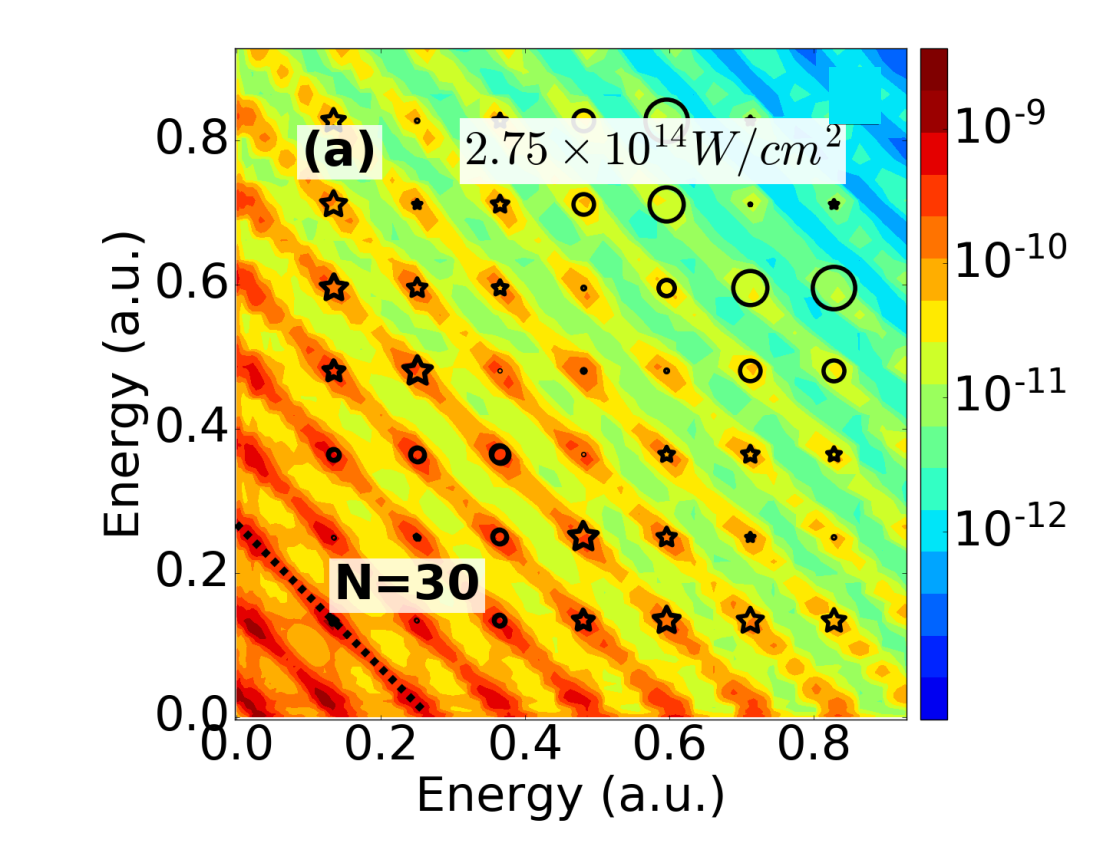}
    \includegraphics[width=0.49\columnwidth, trim=25 25 25 25, clip]{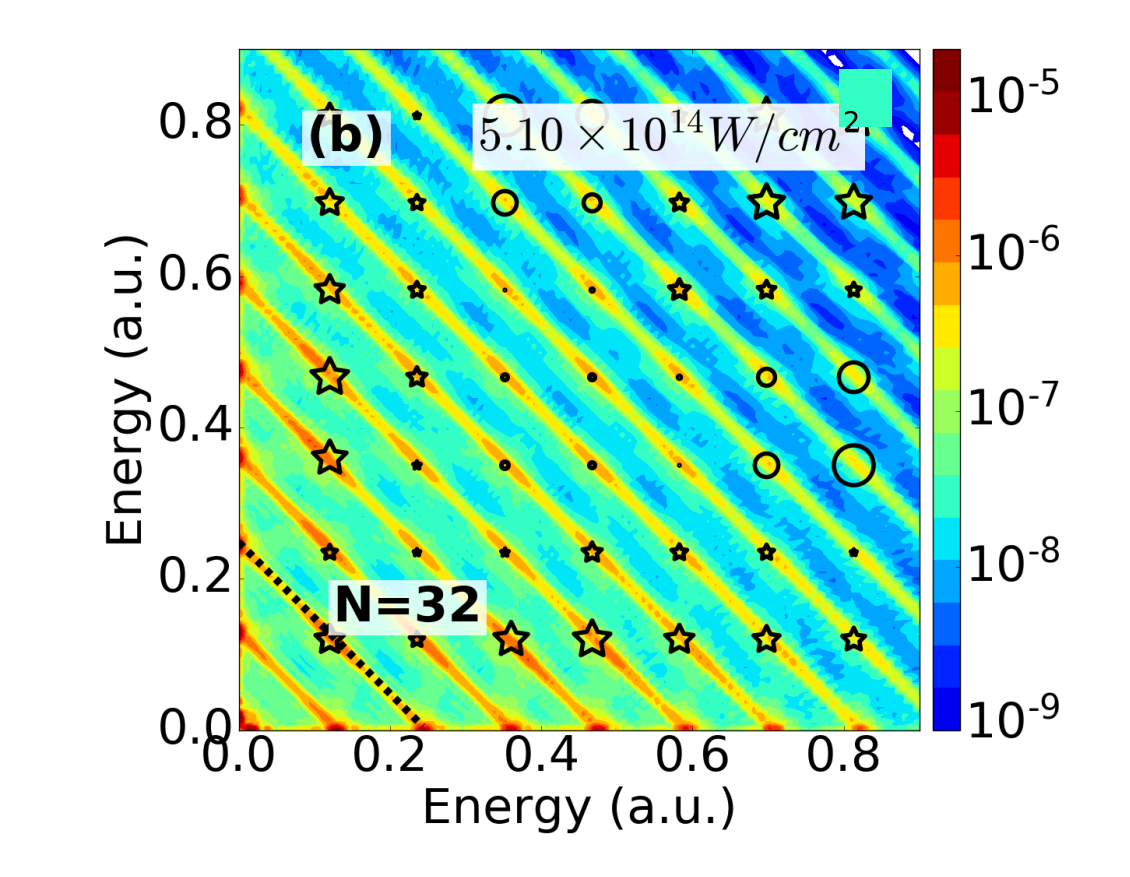}
    \includegraphics[width=0.49\columnwidth, trim=25 25 25 25, clip]{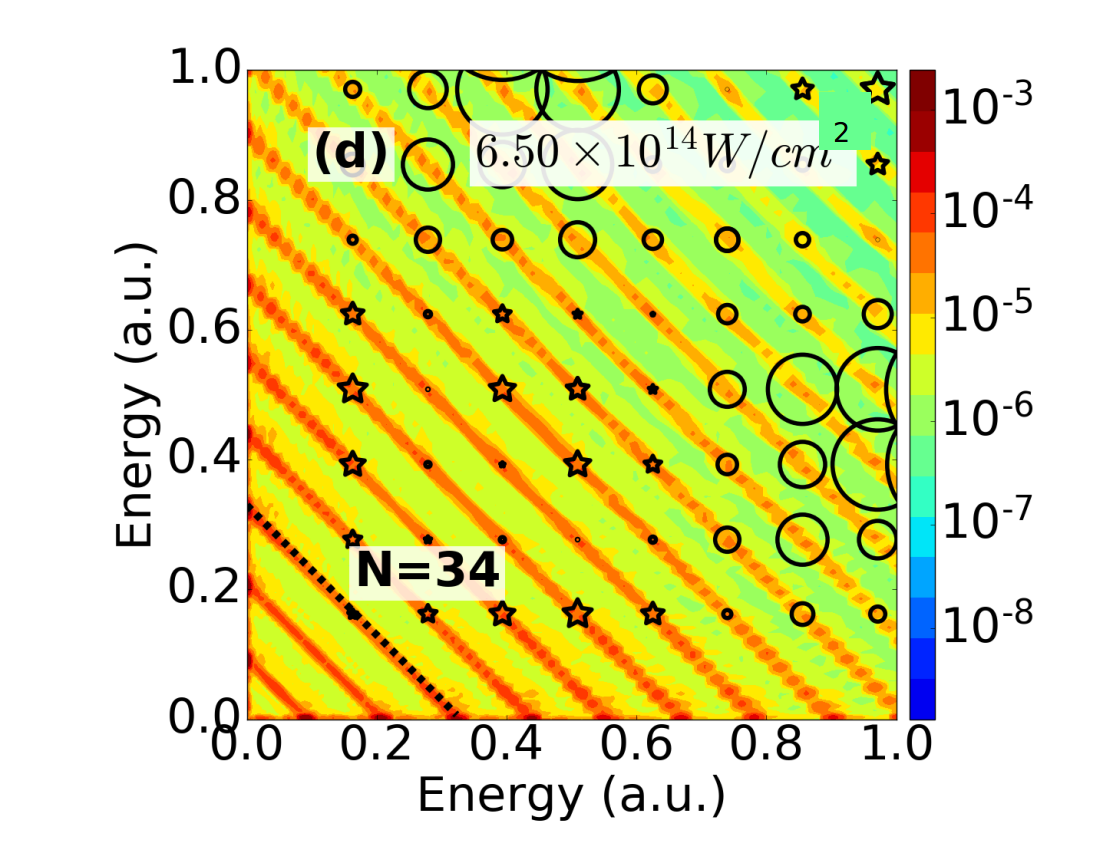}
    \includegraphics[width=0.49\columnwidth, trim=25 25 25 25, clip]{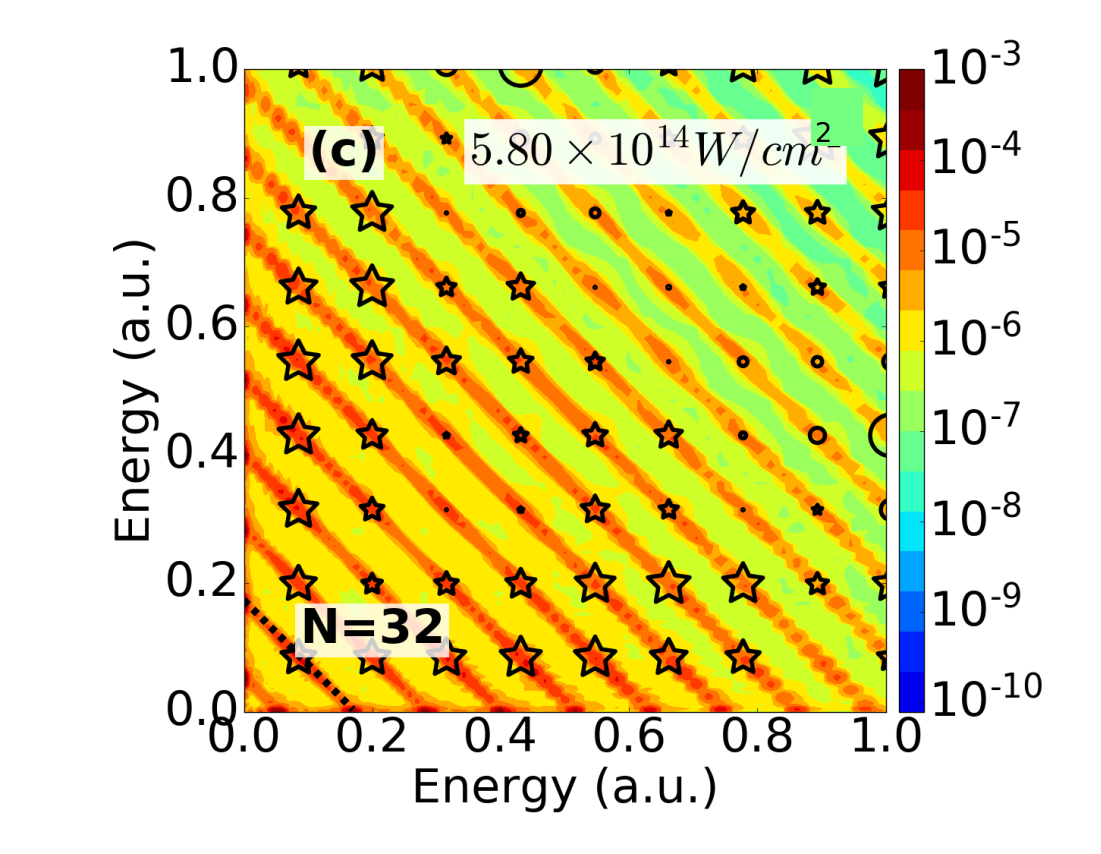}
   \caption{Joint energy distributions  for four different intensities. Stars (circles) indicate anti-correlation, $\Ga_\eta>1$ (correlation, $\Ga_\eta<1$)
   for $\eta=(E_1,E_2)$ and the size of the symbols indicates pronouncements of the effect. 
      (a) and (d) without Freeman resonance, (b) and (c) on resonance, see also Fig.~\ref{fig:intensityScanning}.
      The black lines labeled by $N$ indicate $N$-photon energy peaks in $E_1+E_2$.
      }
    \label{fig:refPoints}
  \end{figure}
  
An overview of the dependence of $\Ga_\eta$ on the photo-electron energies for 4 different intensities is shown in Fig.~\ref{fig:refPoints}.
We see that in general points of non-equal energy sharing are more anti-correlated, $\Ga_\eta>1$. This TDSE result supports
the prediction of preferred anti-correlated emission at non-equal energy sharing \cite{haan08:recollision,ye10:doubleionization}
based on the analysis of classical trajectories. The classical simulations were interpreted by taking into account the modification 
of the classical potential by the simultaneous action of the re-approaching electron and the laser field. In more quantum 
mechanical language this is excitation simultaneous with tunneling and/or over barrier ionization. 
The mechanisms are distinguished from
the conventional idea of RESI (resonant excitation with subsequent ionization) in that excitation and ionization happen within the 
time-frame of a given recollision and therefore the directions of electron emission become 
(anti-)correlated. In contrast, in RESI the
two single ionizations would ultimately occur without narrow correlation in time and leave emission directions largely independent.

A more precise mapping of the mechanisms onto quantum mechanics is difficult: both, the presence of a rather strong field and the brevity of the interaction deprives individual states of their identity. Wavefunctions can, with great success, be associated with 
trajectories at larger distances from the nucleus, but the mapping breaks down as one approaches to within the range of the electrons' de-Broglie wave lengths. Still, the behavior of anti-correlated emission corroborates the essence of Refs.~\cite{haan08:recollision,ye10:doubleionization}: the contribution from ``double-emission upon recollsion'' (DUR) is important, in addition to a possible RESI background.

\begin{figure}[t]
\includegraphics[width=0.9\columnwidth]{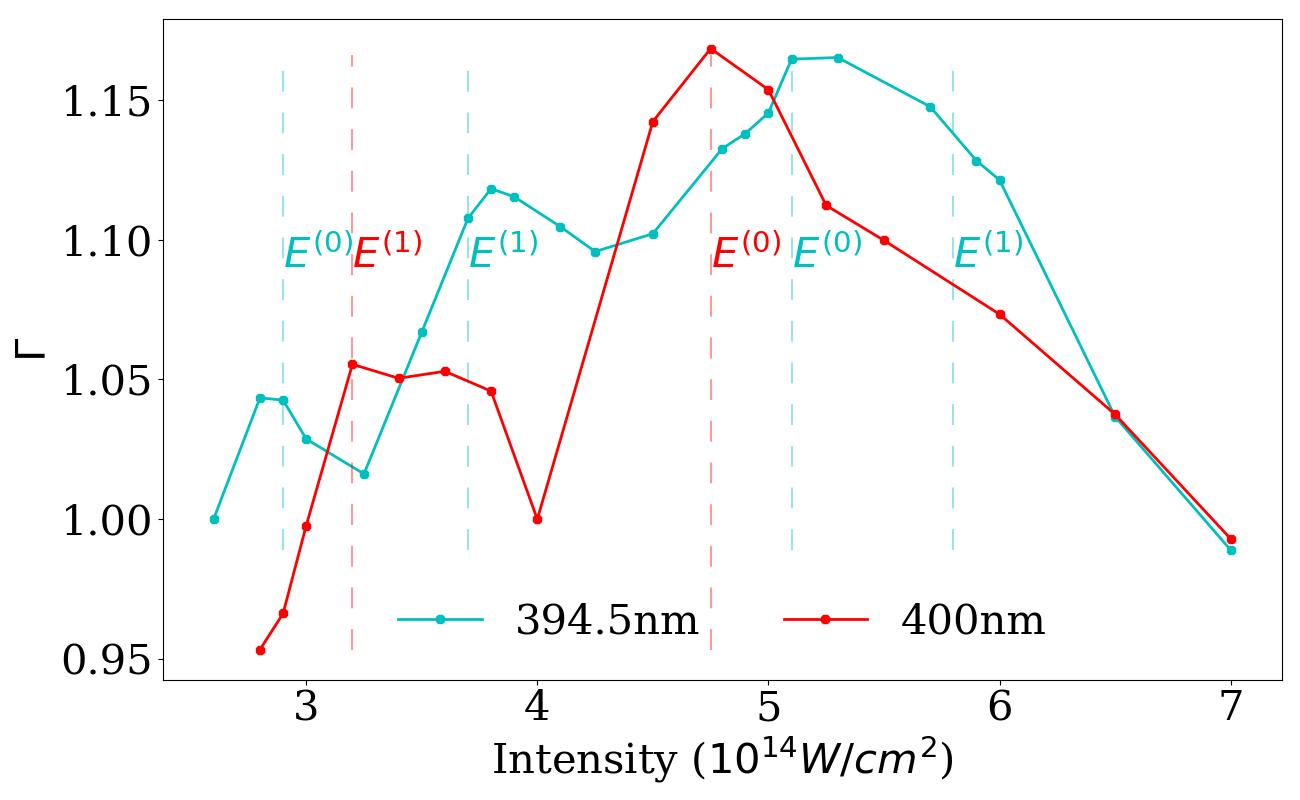}
\caption{Anti-correlation ratio $\Gamma$ as a function of laser intensity with a $\cos^8$ pulse, FWHM=7 \optc\ pulse with $\lambda$ at 394.5nm and 400nm. 
Dashed lines indicate the Freeman resonance positions for the lowest two excited states at peak intensity at the respective 
wave length.
\label{fig:freemanCos8}
}
\label{fig:400nmcos8}
\end{figure}
Fig.~\ref{fig:intensityScanning} was computed with flat-top pulses for better exposure of the mechanism, but Freeman peaks in anti-correlation also appear with the more realistic $\cos^8$ pulse envelope, as shown in Fig.~\ref{fig:freemanCos8}.

Freeman resonances do not appear in classical simulation, as they depend on the quantization of excitation energies. Resonance implies 
in particular that there is a well-defined photon energy and that the process spans several optical periods. In such a mechanism, 
standard multi-photon type excitation is followed double-ionization from the excited state. 
The fact that Freeman resonances enhance anti-correlation indicates that that mechanism is of DUR-type.

\subsection{Joint angular distributions}\label{sec:JADatDI}

\begin{figure}
 \centering 
  \includegraphics[width=\columnwidth]{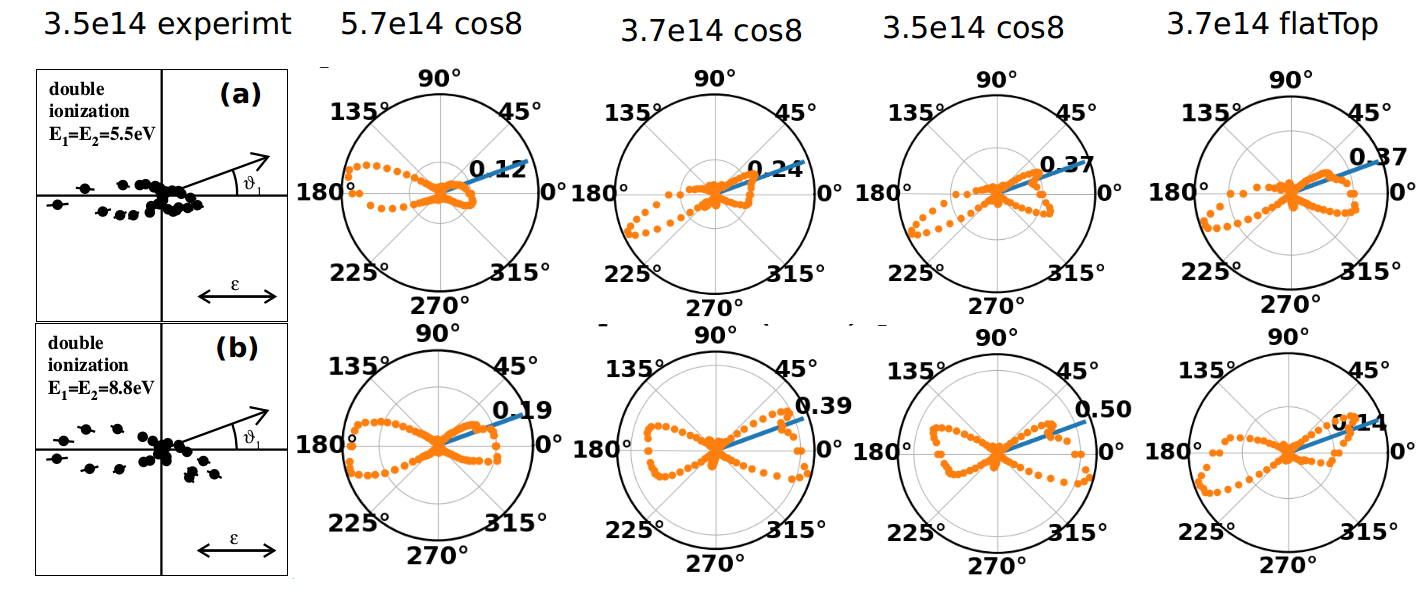}
      \caption{
        JADs from Ref.~\cite{Henrichs2018a} (leftmost) and present simulations at energies $E_1=E_2=5.5\,eV$ (upper row)
        and $=8.8\,eV$ (lower row). Direction of the first electron (blue lines) is fixed at $\th_1=\pi/6$ relative to the polarization axis.
        Intensities and pulse shapes are indicated above the respective columns.
        The distributions are averaged over $\pm 4^\circ$ and normalized to maximal emission 1. A flat-top pulse (last column) does 
        significantly, but not qualitatively change the JAD.
  \label{fig:JADIntensitySensitivity}
  }
\end{figure}

JADs strongly depend on the total energy, the energy sharing between the two electrons, and on the laser 
parameters. 
Fig.~\ref{fig:JADIntensitySensitivity} reproduces two JADs from Ref.~\cite{Henrichs2018a} together with our results.
For illustration we have chosen two points with equal energy sharing $E_1=E_2$ at 5.5 and 8.8\,eV, respectively.
Experiment and simulation agree in showing clear angular anti-correlation.
Near intensity $\inten{3.5}{14}$, the JAD bends into the lower half plane, away from the first emitted electron.
At the higher intensity of $\inten{5.7}{14}$ anti-correlation is less pronounced and shapes are more similar to the experimental ones. Apart from that general qualitative behavior, the spectra vary significantly with the exact pulse shape and intensity.
Because of the high sensitivity to intensities, e.g. comparing 3.5 and $\inten{3.7}{14}$, 
a more detailed comparison of  computed JADs with 
experiment is not possible at this point.

\begin{figure}[b]
  \includegraphics[width=0.8\columnwidth]{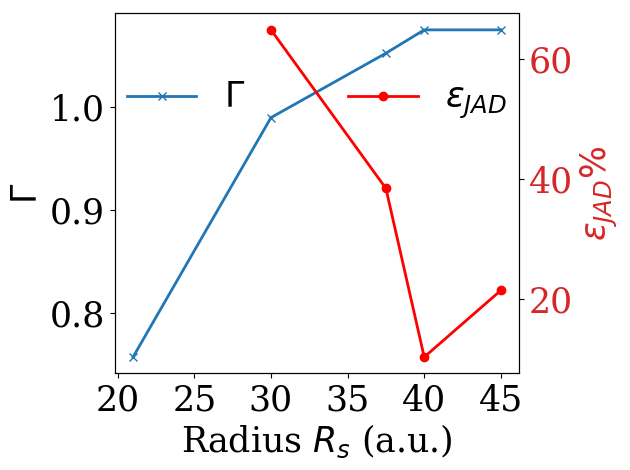}
  \caption{\label{fig:rc-convergence}
  Convergence of $\Ga$ and JADs with the radius $R_s$ of the interaction region. Calculations at $\inten{5}{14}$ and FWHM  of 2 fs.
  $\ep_{JAD}$ is the maximal relative error of the JADs, Eq.~(\ref{eq:epJAD}).
  }
  \end{figure}

By studying the convergence with increasing $R_s$ we see that the bulk of correlation effects originates at distances $\lesssim 30\,au$ from the nucleus. 
Fig.\ref{fig:rc-convergence} shows the convergence of the anti-correlation ratio $\Ga$ and the maximal relative error 
of the energy-integrated angular distributions
\beq\label{eq:epJAD}
\ep_{JAD}=\max_{\th_1,\th_2} \frac{|\si(\th_1,\th_2)-\si^{n}(\th_1,\th_2)|}{\si(\th_1,\th_2)},
\eeq
where $\si^{n}$ refers to results obtained with the next smaller box size.
While $\Ga$ is converged for the purpose of the present argument, convergence of the JADs remains delicate,
but qualitatively correct results may be expected at interaction ranges $R_s\gtrsim40$.

\subsection{Double-emission by short pulses}
\begin{figure}
  \includegraphics[width=0.33\columnwidth,valign=t]{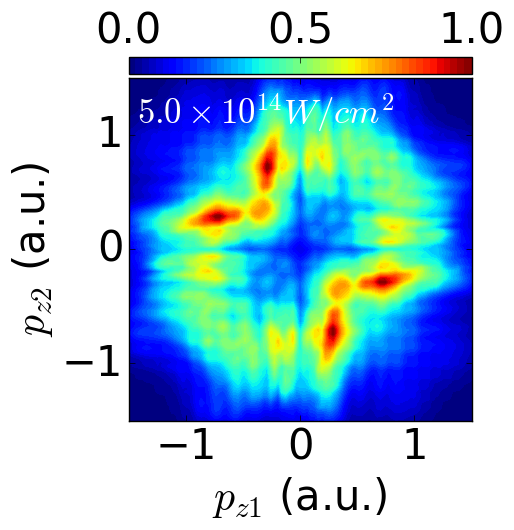}
  \includegraphics[width=0.30\columnwidth,valign=t]{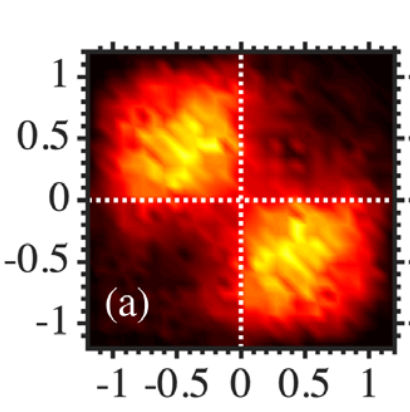}
  \includegraphics[width=0.33\columnwidth,valign=t]{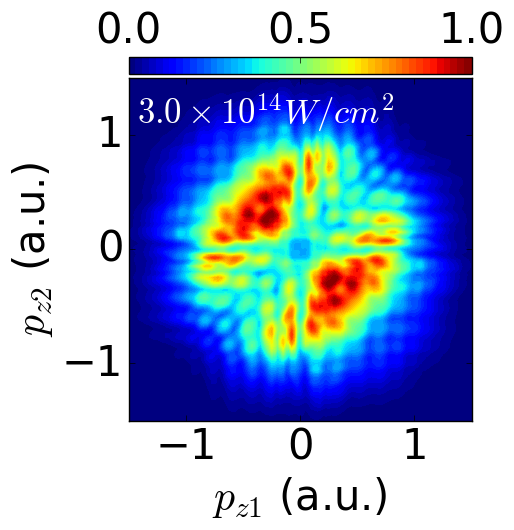}
  \caption{Joint momentum distributions in $p_z$-direction at FWHM 2fs and intensity averaged over carrier-envelope phase. 
  Center: classical trajectory calculation at $\inten{5}{14}$, reproduced from Ref.~\cite{Katsoulis2018}, Fig.1(a). Left: TDSE result for $\inten{5}{14}$, right: TDSE for $\inten{3}{14}$. Densities are normalized to a maximal value of 1.
  \label{fig:pzShort}
  }
\end{figure}

We also investigated double emission by extremely short pulses of 2 fs FWHM with the purpose of identifying a signature
of the "slingshot" mechanism for B2B emission which was proposed in \cite{Katsoulis2018}. In that mechanism, the first electron 
reverts momentum in a close encounter ("slingshot"), while the second electron is emitted with some delay that results in B2B emission. 
Ref.~\cite{Katsoulis2018} reports pronounced B2B emission at the pulse duration of 2 fs and intensity $\inten{5}{14}$ as a
signature of the mechanism. Fig.~\ref{fig:pzShort} compares that classical finding with our TDSE simulations. The result for $\inten{5}{14}$ favors
unequal energy sharing, which is characteristic of a DUR process. In contrast, the $\inten{3}{14}$ result bears great similarity 
with the classical simulation with more weight on equal energy sharing. 

While our finding does not rule the slingshot mechanism at $\inten{5}{14}$, it indicates important double ionization through alternative 
pathways with unequal energy sharing. Note that we use the exact same pulse as in Ref.~\cite{Katsoulis2018}. 

The slingshot 
mechanism may be dominating at the lower intensity. However, attempts to trace the classical motion of the two electrons studying time-dependent
spatial correlations in the quantum wave function failed due to the general difficulty of such a mapping. In addition, we remark
that the very large band width of the 2 fs pulse admits lower order multi-photon ionization, which erodes the quasi-static tunneling picture 
employed for initial ionization in the classical model. Also, by their very construction, classical calculations do not account for effects of the 
quantum mechanical structure of the atom, as for example, the Freeman resonances discussed above.

\section{Conclusions}

The {\it ab initio} quantum mechanical calculations of single- and double-emission confirm the generally 
important role of DUR-type double-ionization, where the second ionization is simultaneous with the recollision, if we accept
the enhancement of B2B emission as a signature of the process. This is supported further by the relatively stronger B2B 
emission at spectral points with large differences between electron energies. 

Comparing with recent experimental results on double emission spectra \cite{Henrichs2018a}, we find good qualitative agreement,
if we allow for an increase of experimental intensities by a factor $\sim 2$. Such an adjustment is suggested by three 
different and largely independent observables: the $He^{++}/He^+$ ratio, the dependence of B2B emission on intensity, and the intensity where the checkerboard pattern in joint-energy distributions fades. 

Unfortunately, the ambiguity of intensity could not be resolved using the single-electron spectra published in \cite{Henrichs2018a}: 
this observable can be computed easily and with great reliability, but we were unable to establish convincing 
agreement at any set of intensities. Again higher than the experimental intensities appear to be favored.

For JADs we can clearly identify the effect of electron repulsion, analogous to what was reported in \cite{Henrichs2018a}.
Comparison with experiment beyond that general level is hampered by the sensitivity of the JADs to intensity, carrier-envelope 
phase, pulse-duration, and exact pulse shape. On the computational side, for reliable convergence of JADs one needs to take into account
the interaction between electrons over large spatial regions $R_s>40$, which inflates tSurff computations to large scale. 

We find that at the given pulse parameters Freeman resonances affect double emission in general and that they disproportionally 
enhance B2B emission. Taking B2B emission as an indicator for a DUR mechanism, this suggests that DI through a Freeman resonance
is primarily DUR.

We finally offer a simple explanation for the checkerboard pattern noted in \cite{Henrichs2018a}, which also appeared in earlier simulations at 800 nm \cite{Zielinski2016}: the modulation at energy-differences of $2\hbar\om$ means that the underlying process involves periodic re-encounters of the two electrons at one-half of the optical period, i.e. multiple recollisions. 

\section*{Acknowledgments}
J.Z. was supported by the DFG Priority Programme 1840, QUTIF. We are grateful for fruitful discussions with G.~Katsoulis, A.~Emmanouilidou K.~Henrichs, and R.~D\"orner.

\bibliography{ratios.bib}
\end{document}